\documentclass[a4paper,11pt]{article}

\usepackage{jinstpub} 

\usepackage{graphicx}
\usepackage{subfig}
\usepackage{multirow}
\usepackage{makecell}
\usepackage{float}         
\usepackage{color, xcolor}
\usepackage{lineno}

\usepackage{amsmath}

\title{\text{Optimization of tracker configuration for the CEPC}}

\author[a,b]{Hao Liang,}
\author[a,b]{Yongfeng Zhu,}
\author[c]{Pei-Zhu Lai,}
\author[a,1]{and Manqi Ruan%
\note{Corresponding author.}}

\affiliation[a]{Institute of High Energy Physics, Beijing 100049, China}
\affiliation[b]{University of Chinese Academy of Sciences, Beijing 100049, China}
\affiliation[c]{ National Central University, No. 300, Zhongda Rd., Taoyuan City 32001, Taiwan}

\emailAdd{ruanmq@ihep.ac.cn}

\abstract{

We investigate the tracker configuration optimization for the Circular Electron Position Collider (CEPC), a proposed Higgs and $Z$ factory. Fixing the construction cost comparable to that of the baseline detector design and considering the benchmark channels ($Z\rightarrow f\bar{f}$, $WW$ fusion with $H\rightarrow f\bar{f}$, $ZH\rightarrow \nu\nu f\bar{f}$, and $t\bar{t}\rightarrow b\bar{b}\mu\nu_\mu ud$) of various operating modes of the CEPC, we obtain the optimal tracker radius that provides the best average resolution of the track momentum or jet energy. The optimal tracker radii for track momentum resolution range from 1.59\,m to 1.73\,m and for jet energy resolution from 1.82\,m to 1.97\,m, depending on the benchmark channels.
Compared to the jets, the tracks prefer a smaller radius and a longer length because the track momentum resolution degrades more significantly than jet energy resolution in the forward region.
The benchmark channel for $Z$-pole prefers a smaller radius and longer length compared to other benchmark channels because the final state particles at the $Z$-pole have a more forward distribution. We also analyze the scaling behavior of the optimal tracker configuration at floating construction cost and observe a weak dependence.

}


\begin{document}
\maketitle
\flushbottom

\section{Introduction}

The Circular Electron Position Collider (CEPC) is a proposed $Z$ and Higgs boson factory~\cite{CEPC_CDR_Acc,CEPC_CDR_Phy}.
It has a main ring with a circumference of 100\,km, which is almost four times the length of the Large Hadron Collider (LHC)~\cite{LHC_TDR_2004}.
Table~\ref{tb:opmode} summarizes its baseline operating scheme and the corresponding boson yields.
As a $Z$-factory, the CEPC will produce nearly one tera of $Z$ bosons at the center-of-mass energy ($E_{\rm cm}$, $\sqrt{s}$) of 91.2\,GeV.
As a Higgs factory, it will produce one million Higgs bosons at $E_{\rm cm}$ of 240\,GeV.
Beyond the baseline operating scheme, $E_{\rm cm}$ of the CEPC can be upgraded to 360\,GeV to study the top quark and the Higgs boson.
Due to the much cleaner collision environment compared to the hadron collider and large integrated luminosity, the CEPC is expected to boost the measurement precision of the properties of the Higgs boson by an order of magnitude on top of the High Luminosity LHC (HL-LHC)~\cite{an2019precision}, to improve the current measurement precision of electroweak by at least an order of magnitude, and to perform extensive studies of flavor physics and QCD.

\begin{table}[!hbtp]
\centering
\caption{The operation scheme of the CEPC, 
including the center-of-mass energy ($\sqrt{s}$),  
the instantaneous luminosity ($\mathcal{L}$), 
the total integrated luminosity ($\int \mathcal{L}$), 
and the event yields.
See~\cite{Gao:2022lew} for more details.}
\label{tb:opmode}

\begin{tabular}{cccccc}
\hline
Operation mode & $\sqrt{s}$ ({\rm GeV}) & Years & \makecell{$\mathcal{L}$ per IP\\ ($10^{34} {\rm cm}^{-2} {\rm s}^{-1}$)} & \makecell{$\int\mathcal{L}$\\ (${\rm ab}^{-1}, 2{\rm IPs}$)} & \makecell{Event\\yields}\\
\hline
$H$        & 240  & 10 & 8.3   & 20  & $4\times 10^6$ \\
$Z$        & 91.2 & 2  & 191.7 & 100 & $3\times 10^{12}$ \\
$W^-W^+$   & 160  & 1  & 26.6  & 6   & $1\times 10^8$ \\ 
$t\bar{t}$ & 360  & 5  & 0.83  & 1   & $5 \times 10^{5}$ \\
\hline
\end{tabular}
\end{table}

\begin{figure}[!hbtp]
\centering
\includegraphics[scale=0.17]{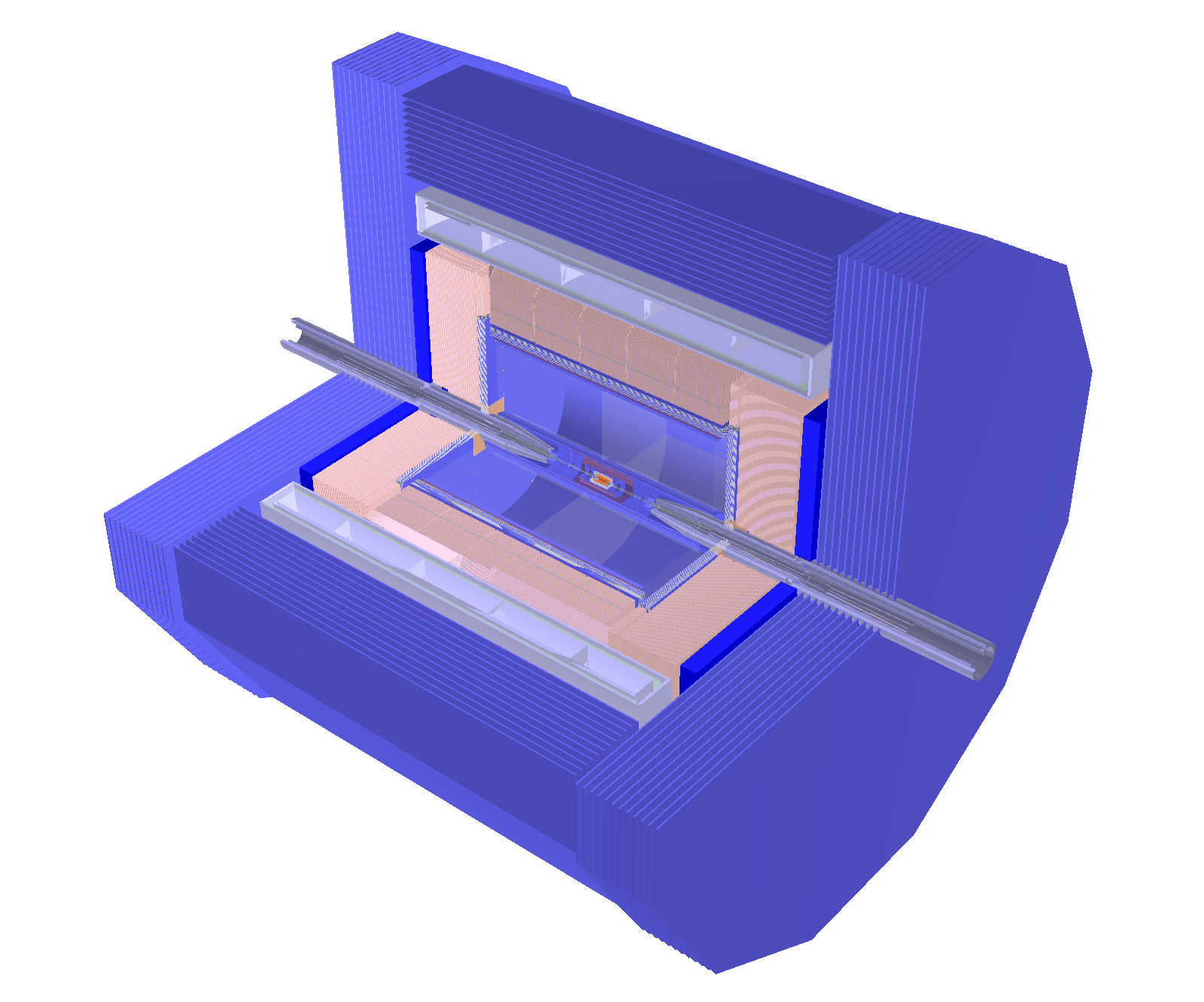}
\caption{
Three-quarters of the CEPC baseline detector~\cite{CEPC_CDR_Phy}.
From inside to outside, it consists of
a vertex detector (red), a silicon tracker,
a time projection chamber (TPC), an electromagnetic calorimeter (ECAL), a hadronic calorimeter (HCAL, pink),
a superconducting solenoid providing an electromagnetic field of 3 Tesla,
and a muon detector.}
\label{fig:detector}
\end{figure}

To fully exploit the enormous physics potential of the CEPC, its detector is required to identify different physics objects efficiently and measure their energy and momentum precisely. The baseline design of the CEPC detector~\cite{CEPC_CDR_Phy} is particle flow algorithm (PFA)~\cite{Ruan_Arbor_2013} oriented, consisting of a high-precision tracking system with a low material budget, a high-granularity calorimeter system, and a 3-Tesla magnet system, as shown in figure~\ref{fig:detector}.
It is targeted at a variety of physics objects~\cite{Ruan_ArborPerf_2018}, including tracks and calorimeter clusters at the subdetector level, final state particles (including leptons, photons, charged and neutral hadrons) at the particle flow level, and high-level composite physics objects such as $\pi^0$, $K^0_S$~\cite{Taifan_2020}, $\Lambda^0$~\cite{Taifan_2020}, $\tau$~\cite{Yu_HiggsTau_2020}, and jets~\cite{Lai_Jet_2021}.

Tracks and jets are the most important physics objects at CEPC.
Because almost 97\% of Higgs events and 70\% of {{$Z$-pole}} events decay into multi-jet final states, and the majority ($\sim$65\%) of the jet energy is carried by charged particles.
The successful reconstruction of tracks not only plays an essential role in the PFA oriented jet reconstruction, it is also critical for any physics measurement involving charged particles in the final state, e.g., Higgs analysis via the $\ell\ell H$ channel, $H\rightarrow\mu\mu/\tau\tau$ measurements~\cite{Yu_HiggsTau_2020}, and most of flavor physics measurements.
In addition, free of the QCD background, hadronic events at CEPC are expected to be reconstructed with high efficiency and purity. The reconstruction of hadronic final states (jets) is thus an advantage of the CEPC compared to the LHC, for example in the analysis of $H\rightarrow bb/cc/gg/WW$ and the measurement of the weak mixing angle in {{$Z$-pole}} operating mode.

\begin{figure}[!hbtp]
\centering
\subfloat[]{\includegraphics[scale=0.137]{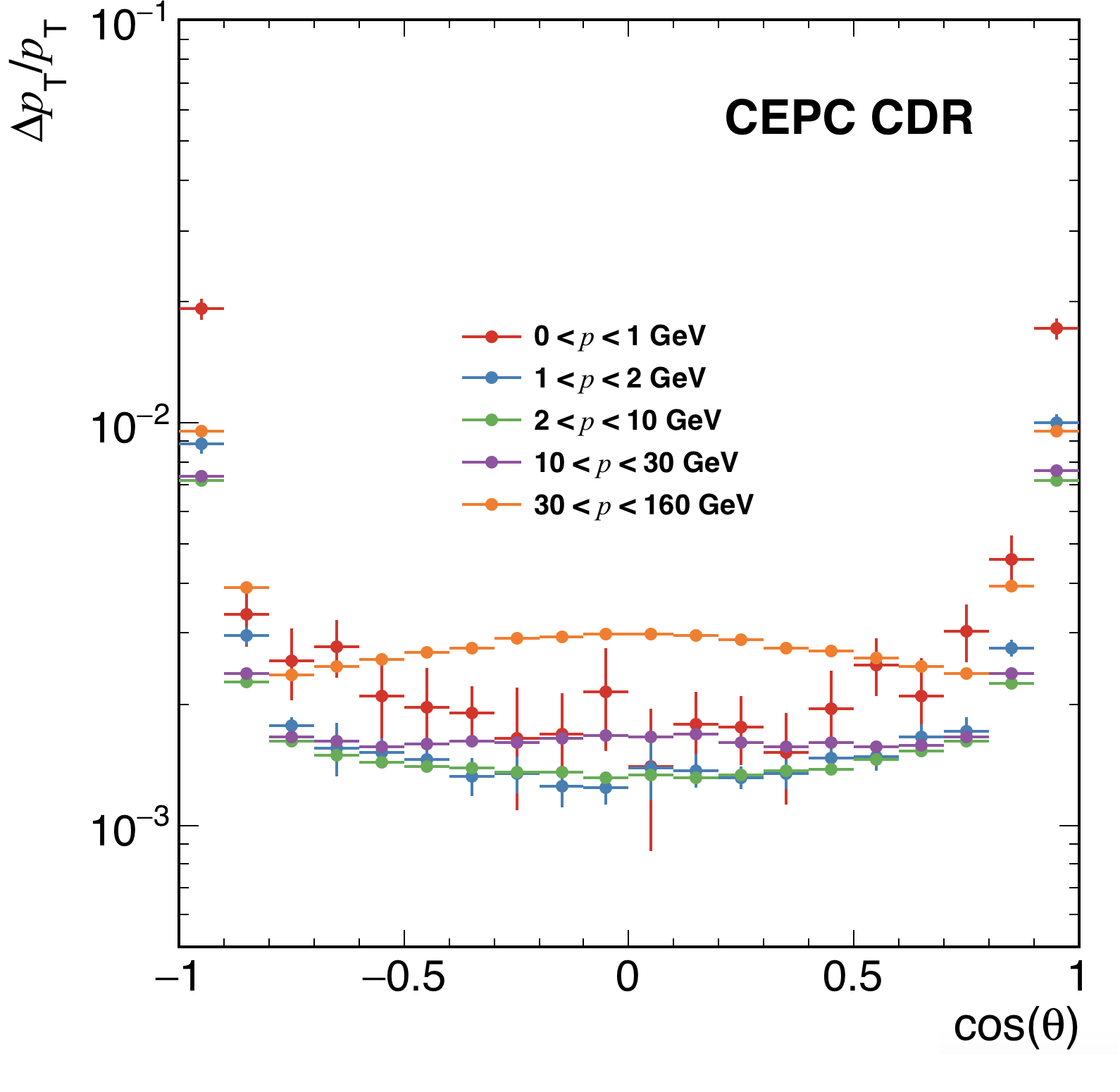} }
\subfloat[]{\includegraphics[scale=0.23]{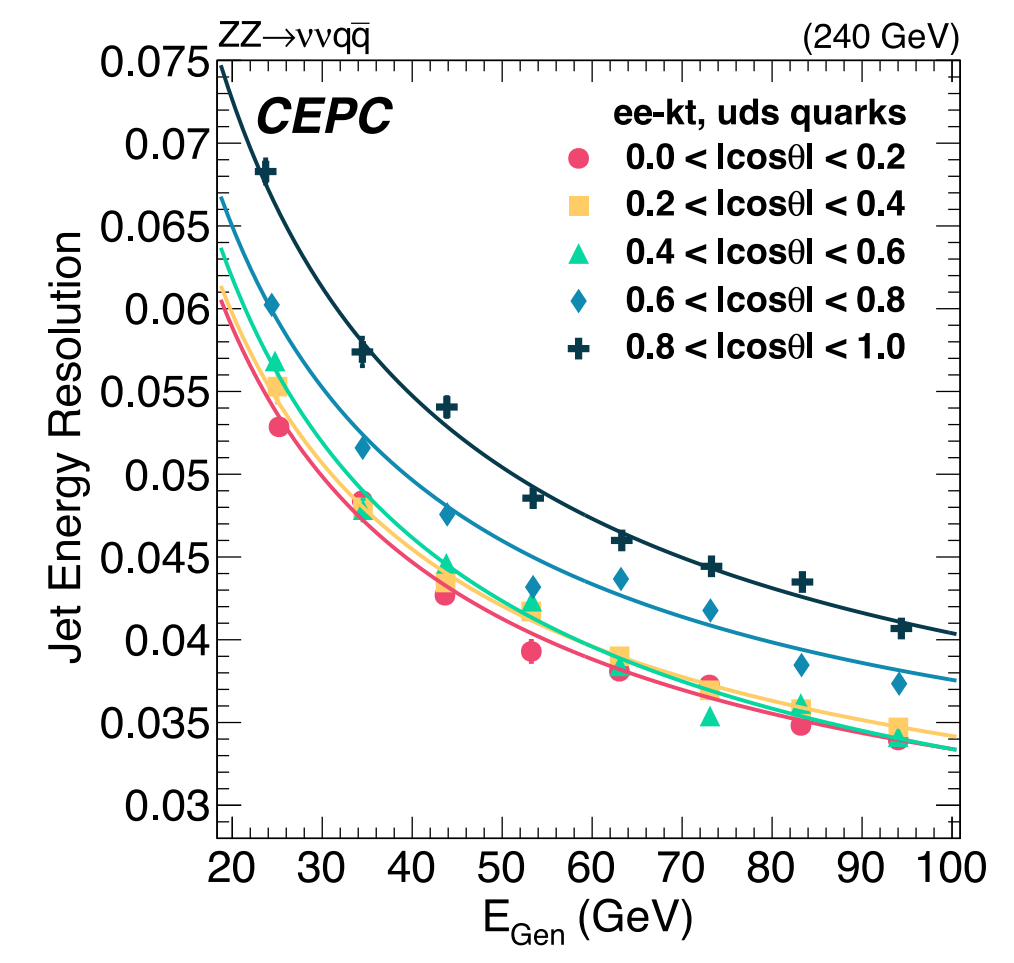} }
\caption{The performance of the CEPC baseline detector for (a) tracks and (b) jets.}
\label{fig:DetectorMuon}
\end{figure}

The CEPC tracker takes a cylinder configuration with a radius of 1.8 meters and a length of 4.7 meters.
Its barrel and endcaps joints at $|\cos(\theta)|$ = 0.8. 
The tracker is highly hermetic with an acceptance characterized by $|\cos(\theta)| < 0.995$, and the acceptance of the calorimeter is even larger.
Figure~\ref{fig:DetectorMuon} shows the measurement resolutions of the tracks and jets with the CEPC baseline detector, 
which provides per-mille level track momentum resolution and 3.5\%--5\% relative jet energy resolution in the barrel region.
However, these resolutions degrade in the endcap region.

\begin{figure}[!hbtp]
\centering
\subfloat[]{\includegraphics[scale=0.20]{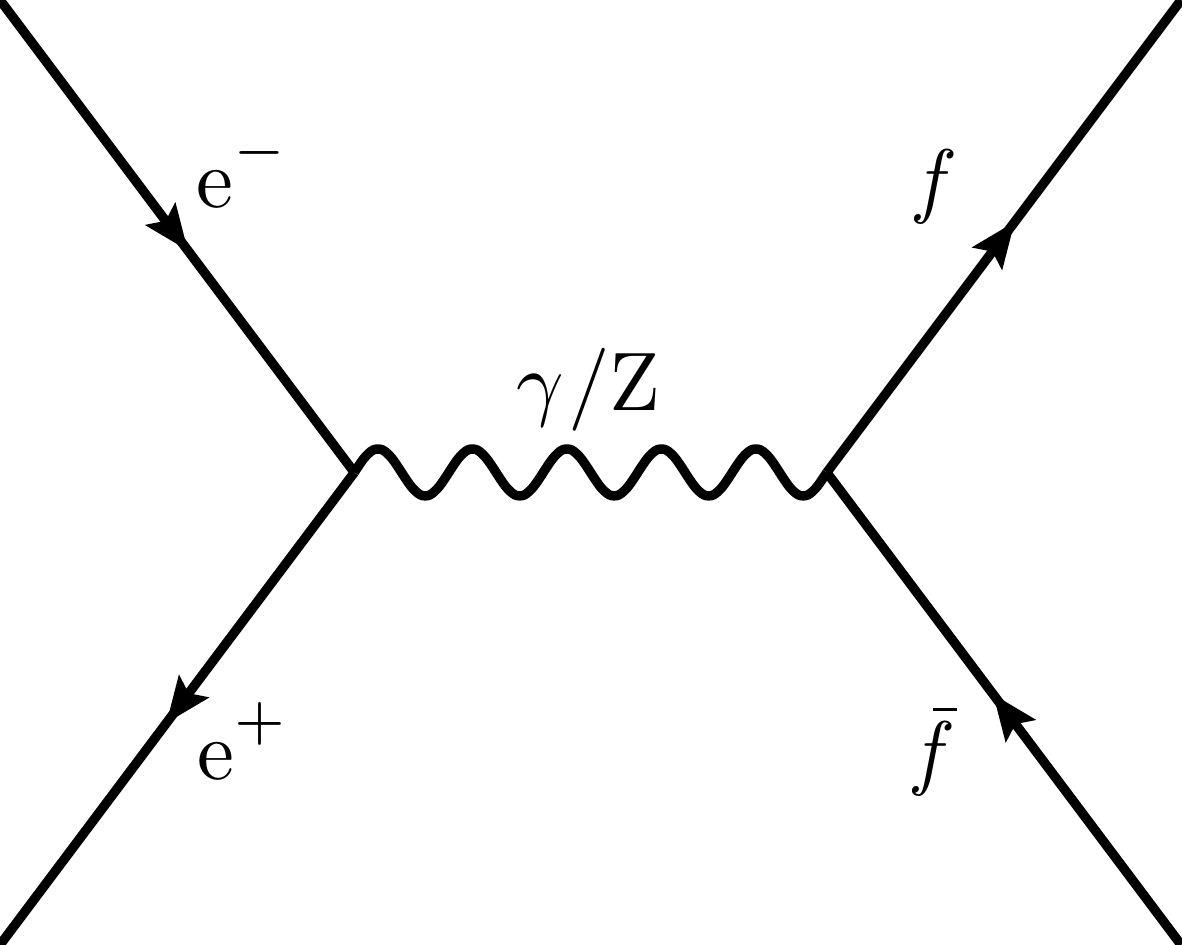} }\quad\quad\quad
\subfloat[]{\includegraphics[scale=0.20]{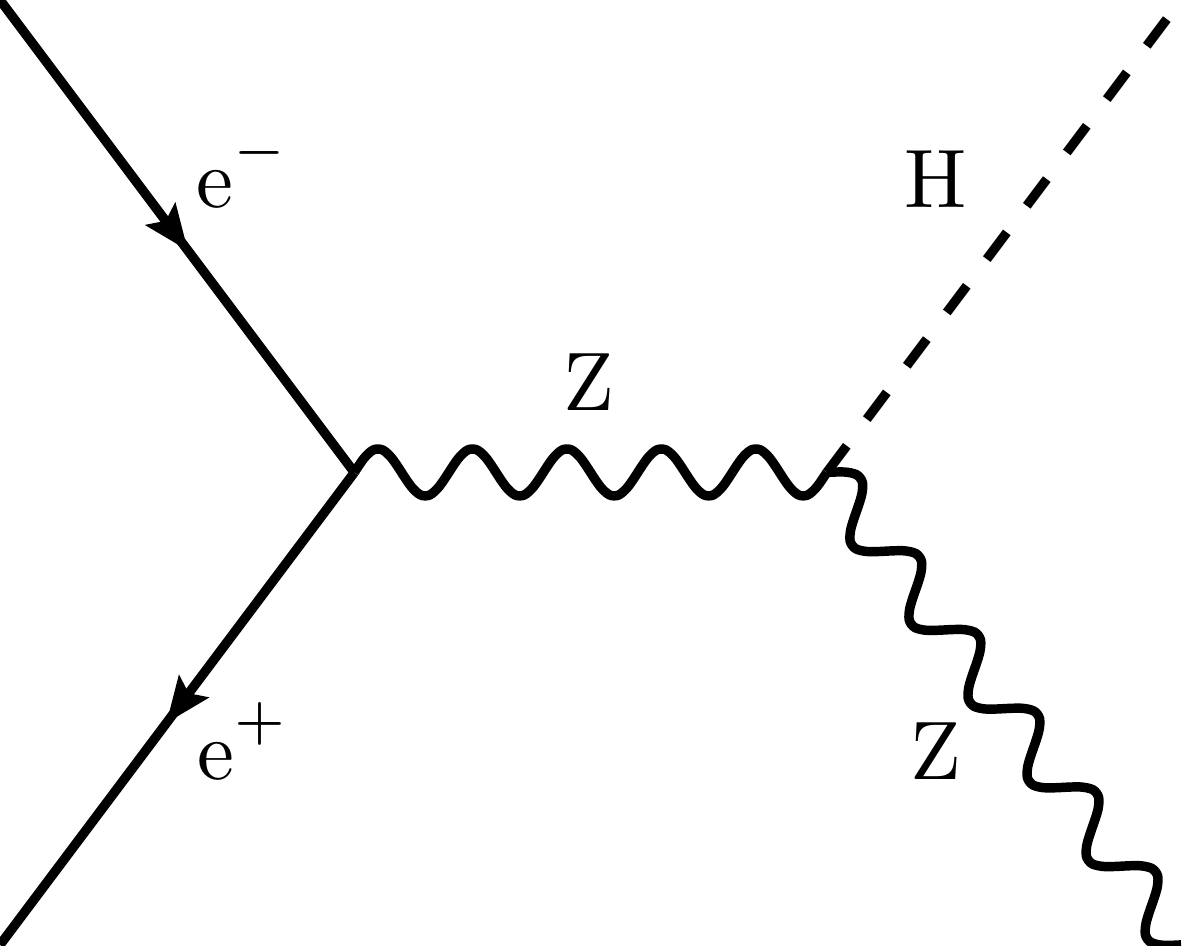}}

\subfloat[]{\includegraphics[scale=0.20]{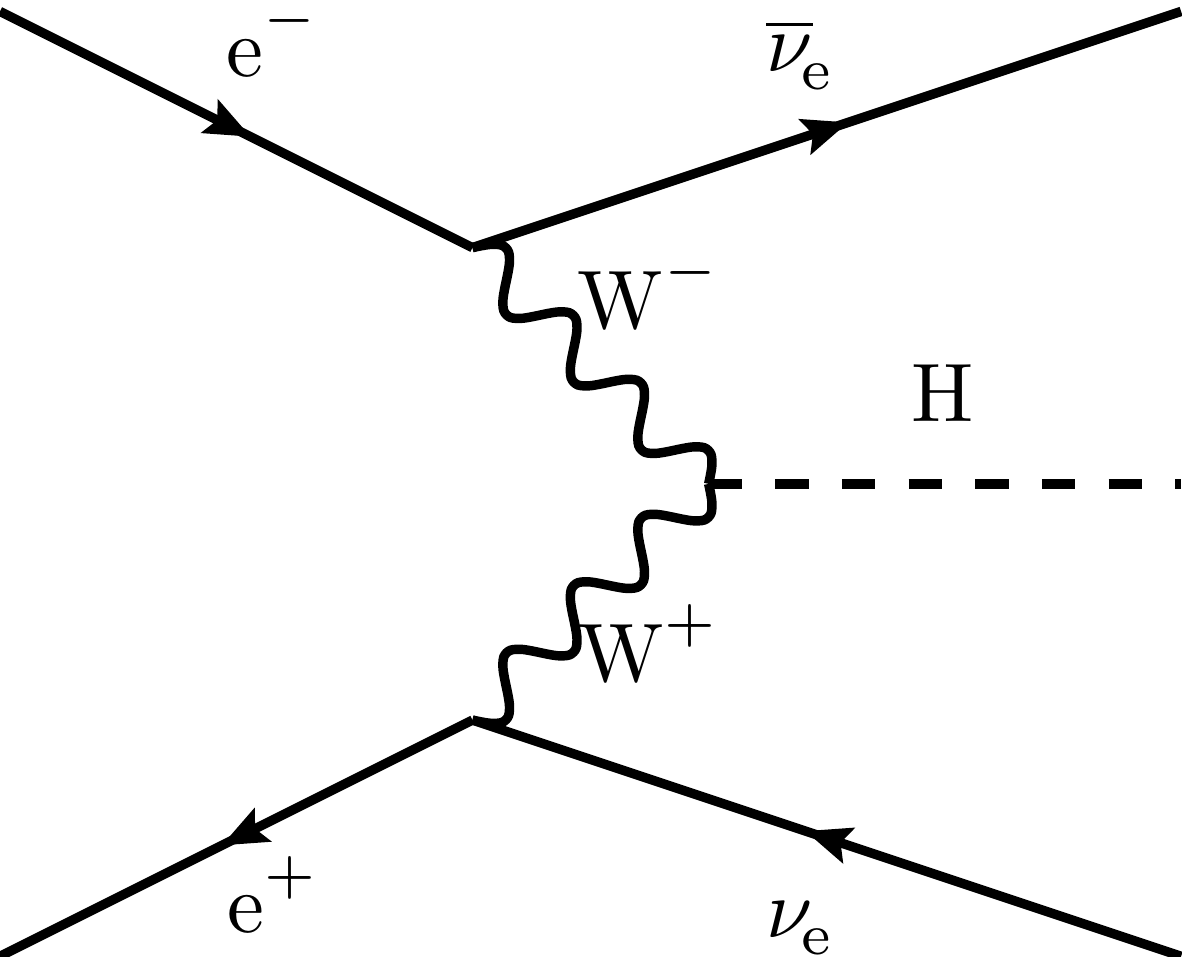} }\quad\quad\quad
\subfloat[]{\includegraphics[scale=0.20]{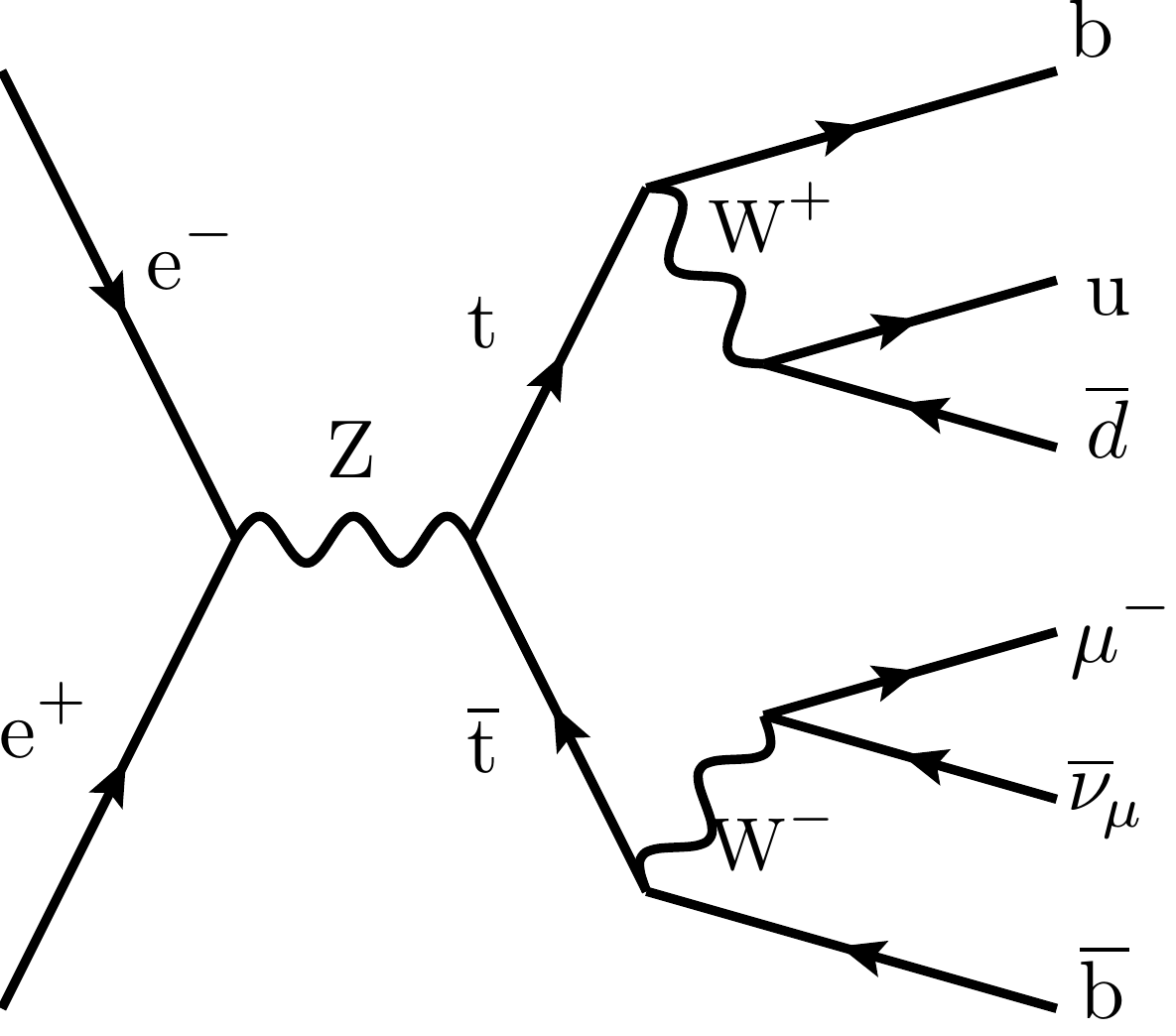} }

\caption{
Feynman diagrams for (a) two-fermion process at $\sqrt{s}$ = 91.2\,GeV,
(b) $ZH$ process at $\sqrt{s}$ = 240\,GeV,
(c) $WW$ fusion process at $\sqrt{s}$ = 360\,GeV,
and (d) $t\bar{t}\rightarrow b\bar{b}\mu\overline{\nu}_\mu u\bar{d}$ at $\sqrt{s}$ = 360\,GeV.}
\label{fig:feyn}
\end{figure}

The reconstruction performance of physics objects depends strongly on the radius and length of the tracker.
The construction cost, which is evaluated by the tracker volume and tracker surface area in this paper, of the detector also depends strongly on these two geometric parameters.
For example, a larger tracker usually has better performance, but also leads to higher construction cost.
In this work, we optimize the tracker configuration, i.e., the tracker radius and the tracker length, with respect to the fixed construction cost.
The tracker configuration optimization is based on several physics benchmark processes, including:
\begin{itemize}
\item[-] $Z\rightarrow\mu^-\mu^+/q\bar{q}$ at $\sqrt{s}$ = 91.2\,GeV,
\item[-] $ZH\rightarrow\nu\nu+\mu^-\mu^+/q\bar{q}$ at $\sqrt{s}$ = 240\,GeV,
\item[-] $WW$ fusion with $H\rightarrow \mu^-\mu^+/q\bar{q}$  at $\sqrt{s}$ = 360\,GeV,
\item[-] {$t\bar{t}\rightarrow b\bar{b}\mu\overline{\nu}_\mu u\bar{d}$ and its charge conjugation at $\sqrt{s}$ = 360\,GeV, both denoted as $t\bar{t}\rightarrow b\bar{b}\mu\nu_\mu u d$.}
\end{itemize}
Their feynman diagrams are shown in figure~\ref{fig:feyn}.
In these four processes, isolated muon tracks are used to optimize the track momentum resolution.
For the optimization of the jet energy resolution, we use the jets hadronized from $q\bar{q}$ in the first three processes and from $b\bar{b}$/$u\bar{d}$ in the $t\bar{t}$ process.
The $WW$ fusion process at $\sqrt{s}$ = 360\,GeV is included because the measurement of its cross section is crucial for the determination of the Higgs width.

This paper is organized as follows.
Section~\ref{sect:sample} describes the methodology of optimization and the resolution modeling of track momentum and jet energy.
In section~\ref{sect:optimization}, we obtain the optimal tracker configuration with the cost fixed to the baseline value.
In section~\ref{sect:variation}, we analyze the impact of the construction cost on the optimal configuration.
Finally, we summarize the conclusions in section~\ref{sect:conc}.

\section{Methodology and resolution modeling}
\label{sect:sample}

The performance of the detector is evaluated as the average measurement resolution of physics objects with differential cross-sections ($\sigma$) of the benchmark channels as weights, which is illustrated as the following formula:
\begin{equation}\
\label{eq:avg_res}
\overline{{\rm Res}}(R,L)
=
\frac{1}{\sigma}
\int
{\rm Res}(\Lambda;R,L)
\frac{{\rm d}\sigma}{{\rm d}\Lambda}
{{\rm d}\Lambda},
\end{equation}
where $\Lambda$, ${\rm Res}(\Lambda;R,L)$, and $\overline{{\rm Res}}$ are the kinematic variable, differential resolution, and average resolution for the physics objects, respectively.
Specifically, Res stands for $\sigma_{p_T}/p_T$ and $\sigma_{E}/E$ for tracks and jets, respectively.
Due to the rotational symmetry of the detector, $\Lambda$ represents $(p_T, \theta)$ and $(E,\theta)$ for tracks and jets, respectively.
In the integral, the acceptance of the detector, $|\cos\theta| < 0.995$ for tracks, is considered.

\begin{figure}[!hbtp]
\centering
 \includegraphics[scale=0.73]{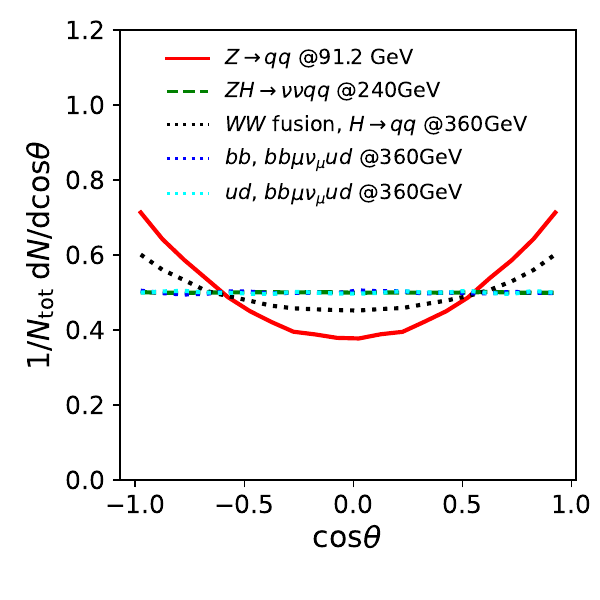}
\caption{The polar angle distributions of   the quarks.
It is a symmetry distribution because we do not distinguish between particles and antiparticles.}
\label{fig:partonDistribution}
\end{figure}

The $\frac{{\rm d}\sigma}{{\rm d}\Lambda}$ is obtained from the \emph{WHIZARD}~\cite{kilian2011whizard} event generator.
Figure~\ref{fig:partonDistribution} shows the polar angle distributions of the quarks in the benchmark channels where the particles and antiparticles are merged.
The muons have almost the same distributions as the quarks in the same process, so the distributions of the muons are not shown in figure~\ref{fig:partonDistribution}.

The quarks at the {{$Z$-pole}} are distributed forward as $1+\cos^2\theta$ in the leading order of electroweak theory~\cite{Workman:2022ynf}.
In the $ZH$ process near threshold, the Higgs bosons are produced almost at rest.
The fermions decaying from the lowly boosted and spin-zero Higgs boson are isotropically distributed.
In the $WW$ fusion process, the fermions from the Higgs decay are significantly forward distributed because of the strongly forward distributed Higgs boson.
In the $tt\rightarrow bb\nu_\mu\mu ud $ channel, near the threshold of top quark production, the fermions have the polar angle distributions of the form $1+S\kappa_{f}\cos\theta_f$~\cite{Jezabek:1988ja, jezabek1994top}, where $S$ denotes the polarization of top quark and $\kappa_f$ is the so-called analysis power for $f$ ($b$, $u$, $d$, $\mu$, and their antiparticles).
$\kappa_f$ have opposite signs for the particle and its antiparticle.
After merging particle and antiparticle, the fermions in the $tt\rightarrow bb\nu_\mu\mu ud $ channel have a flat distribution in $\cos\theta$ .

We model the $\sigma_{p_T}/p_T$ as a function of $p_{T}$ and $\cos\theta$ for the tracks,
and model the $\sigma_{E}/E$ as a function of $E$ and $\cos\theta$ for the jets.
We also parameterize a function by the radius and length of the tracker according to the full simulation results. 
The modeling details for the tracks and jets are presented in sections~\ref{sect:track_trans_mom} and \ref{sect:jet_energy}, respectively.

Finally, we use the construction cost constraint.
The tracker length $L$ can be determined by the tracker radius $R$ once the cost is fixed.
For simplicity, we use $R$ to describe the tracker configuration in the following, and the average resolution can be evaluated as a function of $R$.
We obtain the optimal $R$ as well as $L$, $R/L$, etc., by obtaining the best average resolution.
You can read the results and discussions in section~\ref{sect:optimization}.

\subsection{Track differential transverse momentum resolution}
\label{sect:track_trans_mom}

\begin{figure}[!hbtp]
\centering
\subfloat[]{\includegraphics[scale=0.73]{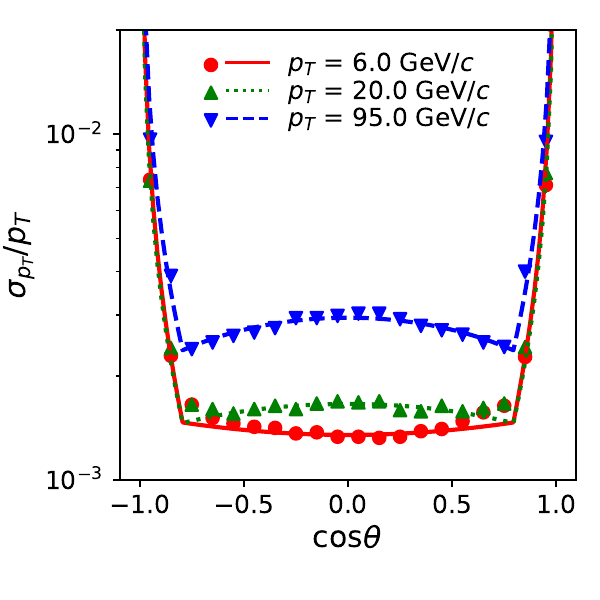} }
\subfloat[]{\includegraphics[scale=0.73]{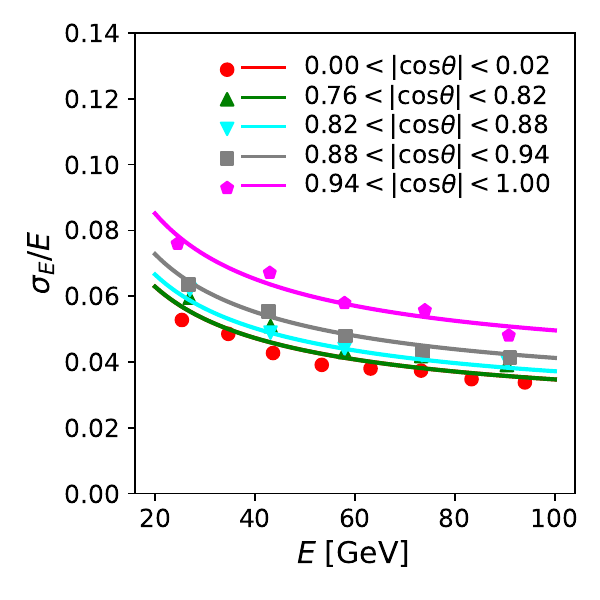} }
\caption{(a) The track transverse momentum resolution as a function of $\cos\theta$;
(b) the jet energy resolution as a function of jet energy.
The markers represent the simulation data with the baseline detector, 
 and the solid lines stand for the parameterization results.}
\label{fig:track_PT_resolution}
\end{figure}

The projection of the track onto the endcap is an arc, and the track transverse momentum resolution is primarily determined by the accuracy of the radius of the arc.
We define the effective radius $r$ as the distance between the position of the outermost detectable track hit and the $z$-axis. 
For a particle with an elementary charge and a transverse momentum of $p_T$ flying in a magnetic field of 3 T, the radius of the arc is approximately 1.1$\frac{p_T}{{\rm GeV}/c}\text{ m}$.
Typically, the $p_T$ of an isolated track is several GeV$/c$, so the radius of the arc is several times larger than the effective radius $r$.
The effective radius is approximated by eq.~\eqref{eq:r}:
\begin{equation}
\label{eq:r}
\quad
r = R
\left\{
\begin{aligned}
&1                            &&  |\cos\theta|\le \cos\theta^c, \\
&\tan\theta/\tan\theta^c && |\cos\theta| > \cos\theta^c,
\end{aligned}
\right.
\end{equation}
where $\theta^c$ is the polar  angle corresponding to the intersections of the barrel and endcaps.
The larger the arc length for a given $p_{T}$ value, the better the track momentum resolution can be.
This explains the degradation of $\sigma_{p_T}$/$p_T$ when the particle hits the endcaps.
In the first-order approximation, $\sigma_{p_T}$/$p_T$ is proportional to $r^{-2}$~\cite{karimaki1998explicit} and can be parameterized as:
\begin{equation}
\label{eq:trackTMR}
\begin{split}
\sigma_{p_T}/p_T  &= 
\frac{r^{-2}}{R_0^{-2}}(c_0 + c_1 \cos^2 \theta),\\
\ln c_0 &= a_0 + b_0 \ln p_T + d_0 \ln^2 p_T,\\
 c_1 &= a_1 + b_1 \ln p_T + d_1 \ln^2 p_T.
\end{split}
\end{equation}
We use the terms in the parentheses in eq.~\eqref{eq:trackTMR} to address the remaining resolution dependence on $p_T$ and  $\cos\theta$.
The parameters $a_i$, $b_i$, and $d_i$ are obtained by fitting to the full simulation results~\cite{CEPC_CDR_Phy} for the baseline.
The fitted results are shown in figure~\ref{fig:track_PT_resolution}, which shows good agreement between the modeling and simulation results.

\subsection{Jet differential energy resolution}
\label{sect:jet_energy}

The concept of the effective radius $r$ in eq.~\eqref{eq:r} is also used in modeling jet energy resolution (JER).
The jet structure makes the parameterization of JER is more difficult than that of the track momentum resolution.
The variation of $r$ affects not only the accuracy of the momentum measurements for each track,  but also the particle separation power required for the PFA. 
Therefore, we parameterize the JER using the following formula:
\begin{equation}
\label{eq:BvsE}
\begin{split}
\sigma_E/E &= 
\sqrt{
c_1(E) + \frac{c_2(E)}{r_0 + r}
}\\
c_i &=a_i + b_i/E, i = 1,2
\end{split}
\end{equation}
The $r_0$ is used to ensure that the $\sigma_E/E$ is well defined at $r=0$.
We obtain the parameters $a_i$, $b_i$ ($i=1,2$), and $r_0$ by fitting the formula to the full simulation results obtained using the same method as in the reference~\cite{Lai_Jet_2021}.
The results of the fitting  are shown in figure~\ref{fig:track_PT_resolution}.

\section{Optimization of the tracker configuration}
\label{sect:optimization}

\begin{figure}[!hbtp]
\centering
\includegraphics[scale=0.73]{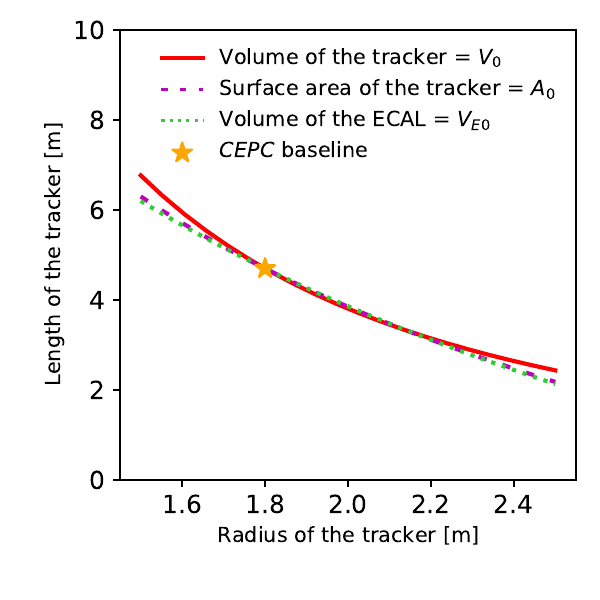}
\caption{The tracker length $L$ as a function of the tracker radius $R$ when fixing the cost to the baseline. 
For the calculation of volume and surface area, the tracker is treated as a cylinder with a radius $R$ and a length $L$.  The calorimeter is treated as a hollow cylinder with an inner radius $R$, an inner length $L$, and a thickness of one meter. For the CEPC baseline design, the volume of tracker is $V_0 = 47.84\,$m$^3$, and the surface area of the tracker is $A_0 = 73.51\,$m$^2$.}
\label{fig:ZvsR}
\end{figure}

In this section, we research the optimal configuration with the construction budget fixed to the baseline.
Figure~\ref{fig:ZvsR} shows the relationship of the tracker length $L$ versus the radius $R$ with the fixed cost.
In this work, the cost of the detector can be estimated by the volume of the tracker, the surface area of the tracker, and the volume of the entire calorimeter system.
In principle, a more realistic estimator could be a linear combination of these estimators with different weights.
However, since the weights are not known, we will optimize the tracker configuration with respect to each cost estimator.
The optimal configuration with respect to the combination of these estimators should be in the middle of the optimal configurations for each cost estimator.
For the calorimeter volume estimator and the tracker surface area estimator, the dependencies of $L$ on $R$ are similar, so we discuss only the tracker volume estimator and the tracker surface area estimator below.

\begin{figure}[!hbtp]
\centering

\subfloat[]{\includegraphics[scale=0.73]{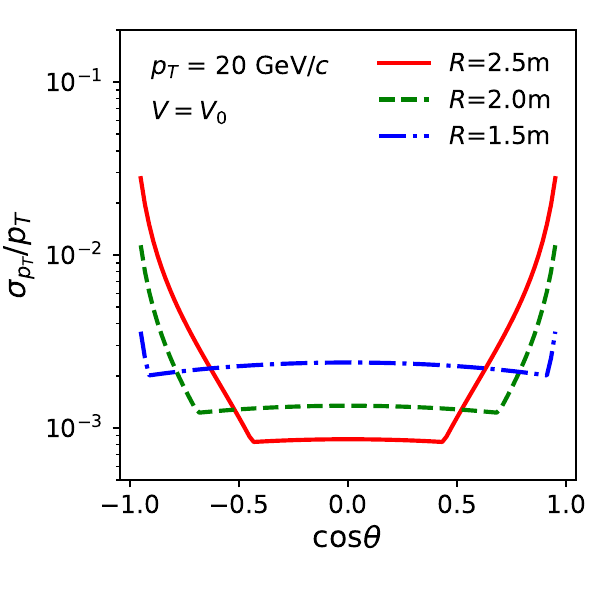} }
\subfloat[]{\includegraphics[scale=0.73]{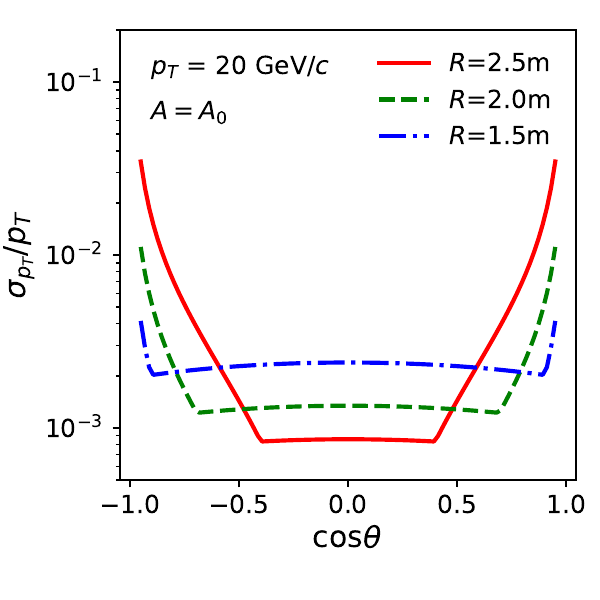} }

\subfloat[]{\includegraphics[scale=0.73]{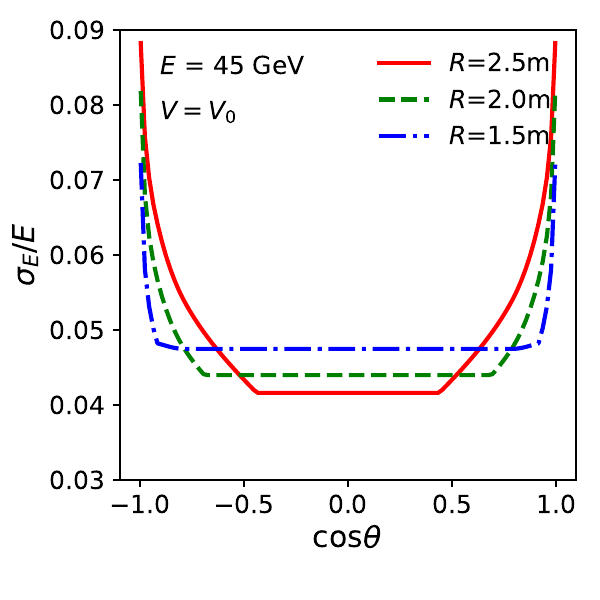} }
\subfloat[]{\includegraphics[scale=0.73]{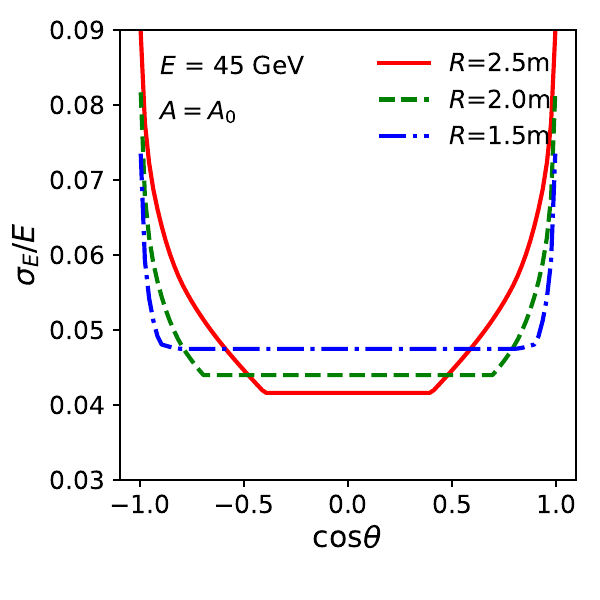} }

\caption{
Top panel: the transverse momentum resolution as a function of $\cos\theta$
for a track with a momentum of 20 GeV/$c$.
Either (a) the tracker volume or (b) the tracker surface area is fixed to the baseline.
Bottom panel: the jet energy resolution as a function of $\cos\theta$ for a jet with an energy of 45 GeV.
Either (c) the tracker volume or (d) the tracker surface area is fixed to the baseline.
}
\label{PT_resolution_Fix}
\end{figure}

Using eq.~\eqref{eq:r}--\eqref{eq:trackTMR}, the track transverse momentum resolution is predicted as a function of $p_T$ and $\cos\theta$.
For a given track momentum, the track transverse momentum resolution as a function of $\cos\theta$ is shown in figure~\ref{PT_resolution_Fix}.
All curves have a U-shaped valley.
The wide bottom corresponds to the tracks hitting the barrel,
and the two sides are the tracks hitting the endcaps.
The closer the track is to the $z$-axis, the smaller the effective radius becomes, resulting in a rapid degradation of the track momentum resolution.

Similarly, with eq.~\eqref{eq:BvsE}, the jet energy resolution is predicted as a function of $\cos\theta$ and $E$.
For a given jet energy, the jet energy resolution as a function of $\cos\theta$ is shown in figure~\ref{PT_resolution_Fix}.
The figure shows a similar pattern, a U-shaped valley, to that of the track.
However, the dependence of JER on the polar angle is much weaker than that of the track momentum resolution,
since the measurement of the jets can always rely on the ECAL, especially in the forward region of the detector.
The performance of the jets in the endcaps is up to two times worse than that in the barrel.

\begin{figure}[!hbtp]
\centering
\subfloat[]{\includegraphics[scale=0.73]{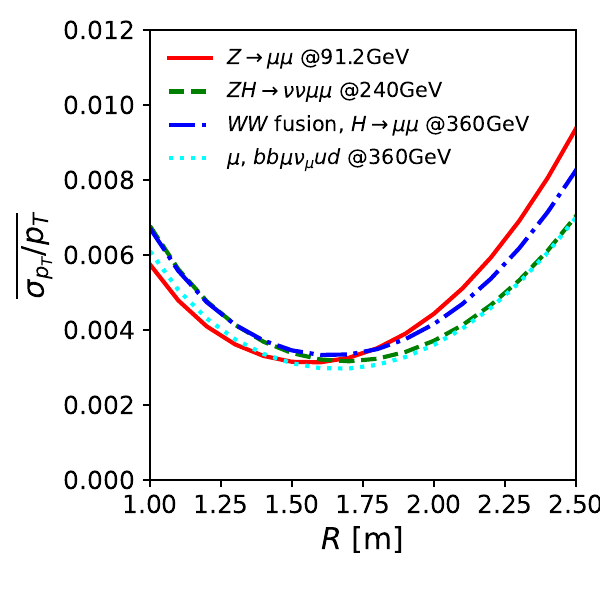} }
\subfloat[]{\includegraphics[scale=0.73]{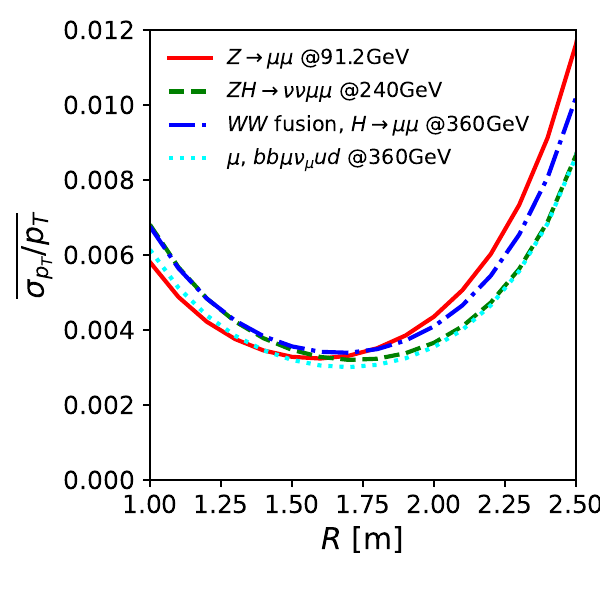} }

\subfloat[]{\includegraphics[scale=0.73]{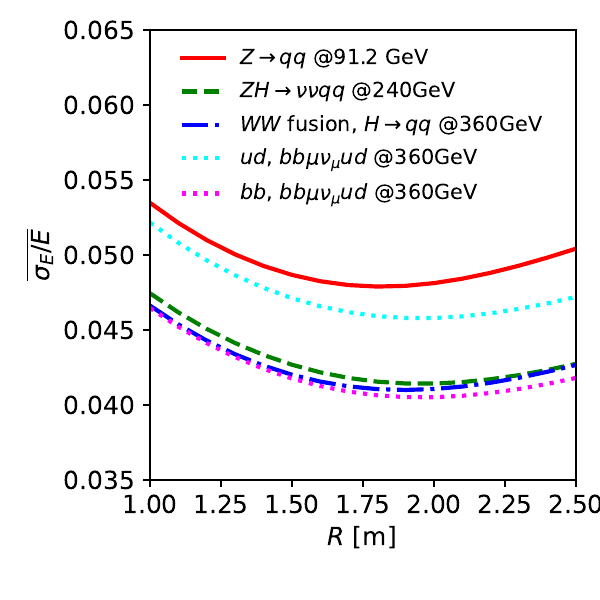} }
\subfloat[]{\includegraphics[scale=0.73]{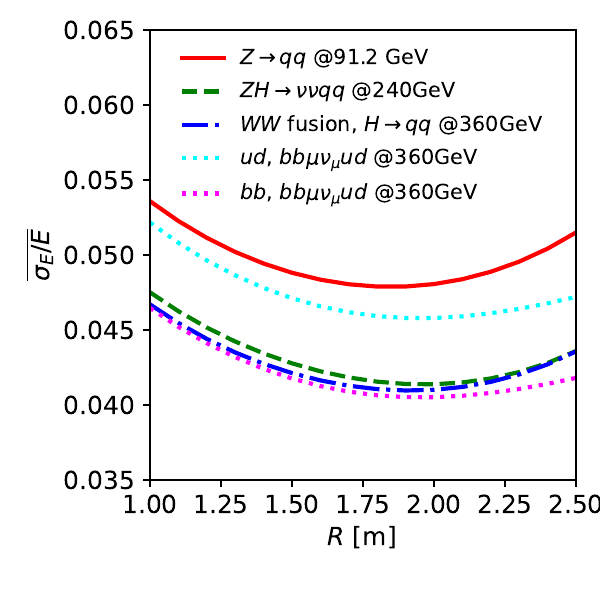} }
\caption{
Top panel: the average transverse momentum resolution as a function of $R$.
Either (a) the tracker volume or (b) the tracker surface area is fixed to the baseline.
Bottom panel: the average jet energy resolution as a function of $R$.
Either (c) the tracker volume or (d) the tracker surface area is fixed to the baseline.
}
\label{fig:MuonJetResults}
\end{figure}

As the radius $R$ increases (and the length $L$ decreases respect to  the fixed cost), the measurement performance of the tracks and jets gets better in the barrel and gets worse in the endcaps, as shown in figure~\ref{PT_resolution_Fix}.
There is an optimal $R$ to achieve the best average resolution over the phase space of physics objects.
The average track momentum resolution and the average JER are calculated as functions of the tracker radii, as shown in figure~\ref{fig:MuonJetResults}.
Table~\ref{tb:results} summarizes the optimization results for each channel and cost constraint.

\def\volume{volume}
\def\area{surface area}

\begin{table}[!hbtp]
\centering
\caption{The optimal tracker configurations and the corresponding average resolutions. At the CEPC baseline, the volume of tracker is $ 47.84\,$m$^3$, and the surface area of the tracker is $73.51\,$m$^2$.}
\label{tb:results}

\begin{tabular}{cccccc}
\hline
Benchmark & \makecell{Cost estimator} & \makecell{Optimal\\$R$  (m)} & \makecell{Optimal \\$L$ (m)} & \makecell{Optimal\\$R/L$} 
&  \makecell{Optimal \\ resolution (\%)} \\
\hline

\multirow{2}*{\makecell{$Z\rightarrow\mu^-\mu^+$\\$\sqrt{s} =91.2$ GeV  } }

& \volume& 1.59 & 5.78 & 0.275 & 0.324 \\
& \area & 1.56 & 6.26 & 0.249 & 0.313 \\\hline

\multirow{2}*{\makecell{$Z\rightarrow q\bar{q}$ \\ $\sqrt{s} =91.2$ GeV } }

& \volume & 1.85 & 4.48 & 0.412 & 4.788 \\
& \area & 1.82 & 4.62 & 0.393 & 4.790 \\
\hline

\multirow{2}*{\makecell{$ZH\rightarrow \nu\nu \mu^-\mu^+$ \\ $\sqrt{s} =240$ GeV  } }
& \volume & 1.72 & 5.07 & 0.339 & 0.319 \\
& \area & 1.69 & 5.32 & 0.318 & 0.315 \\\hline

\multirow{2}*{\makecell{$ZH\rightarrow \nu\nu q\bar{q}$ \\ $\sqrt{s} =240$ GeV } }
& \volume & 1.97 & 3.98 & 0.494 & 4.265 \\
& \area & 1.96 & 3.98 & 0.492 & 4.270 \\
\hline

\multirow{2}*{\makecell{$WW$ fusion, $H\rightarrow \mu^-\mu^+$ \\ $\sqrt{s} =360$ GeV } }
& \volume & 1.67 & 5.33 & 0.313 & 0.339 \\
& \area & 1.64 & 5.66 & 0.290 & 0.332 \\
\hline

\multirow{2}*{\makecell{$WW$ fusion, $H\rightarrow q\bar{q}$ \\ $\sqrt{s} =360$ GeV } }
& \volume & 1.92 & 4.19 & 0.458 & 4.097 \\
& \area & 1.90 & 4.24 & 0.448 & 4.100 \\
\hline

\multirow{2}*{\makecell{$\mu^{\pm}$, ${ b\bar{b} \mu \nu_\mu ud }$ \\ $\sqrt{s} =360$ GeV } }
& \volume & 1.69 & 5.24 & 0.322 & 0.302 \\
& \area & 1.66 & 5.55 & 0.298 & 0.297 \\
\hline

\multirow{2}*{\makecell{$ud$,  ${ b\bar{b} \mu \nu_\mu ud }$  \\ $\sqrt{s} =360$ GeV } }
& \volume & 1.96 & 4.00 & 0.491 & 4.580 \\
& \area & 1.95 & 4.00 & 0.488 & 4.585 \\
\hline

\multirow{2}*{\makecell{$b\bar{b}$,  ${ b\bar{b} \mu \nu_\mu ud }$  \\ $\sqrt{s} =360$ GeV } }
& \volume & 1.97 & 3.98 & 0.495 & 4.046 \\
& \area & 1.96 & 3.97 & 0.493 & 4.050 \\
\hline

\end{tabular}
\end{table}

\begin{table}[!hbtp]
\centering
\caption{ The performance degradations for different tracker radii compared to the optimal resolution of each benchmark channel.
The box shows the minimum number of each row. }
\label{fig:compare}
\begin{tabular}{ccccccccccc}
\hline
\multirow{2}{*}{Benchmark} & 
\multirow{2}{*}{Cost estimator} &
\multicolumn{8}{c}{ Degradations (\%) vs. radii (m) }\\
 & & 1.5 & 1.6 & 1.7 & 1.8 & 1.9 & 2.0 & 2.1 & 2.2\\
\hline

\multirow{2}*{\makecell{$Z\rightarrow \mu^- \mu^+$\\$\sqrt{s} =91.2$ GeV  } }

&\volume & 0.8 & \boxed{0.3} & 4.2 & 12.4 & 24.9 & 41.8 & 63.3 & 89.6\\
&\area & 1.4 & \boxed{0.0} & 2.3 & 8.5 & 19.0 & 34.6 & 56.3 & 86.1\\
\hline

\multirow{2}*{\makecell{$Z\rightarrow q\bar{q}$ \\ $\sqrt{s} =91.2$ GeV } }

&\volume & 1.6 & 0.7 & 0.2 & \boxed{0.0} & 0.1 & 0.5 & 1.1 & 1.9\\
&\area & 2.0 & 1.0 & 0.4 & \boxed{0.0} & 0.0 & 0.4 & 1.1 & 2.1\\
\hline

\multirow{2}*{\makecell{$ZH\rightarrow \nu\nu \mu^- \mu^+$ \\ $\sqrt{s} =240$ GeV  } }
&\volume & 6.9 & 1.6 & \boxed{0.0} & 2.2 & 7.9 & 17.3 & 30.4 & 47.4\\
&\area & 8.5 & 2.5 & \boxed{0.1} & 1.1 & 5.7 & 14.4 & 28.0 & 47.9\\
\hline

\multirow{2}*{\makecell{$ZH\rightarrow \nu\nu q\bar{q}$ \\ $\sqrt{s} =240$ GeV } }
&\volume & 3.1 & 1.8 & 0.9 & 0.3 & 0.0 & \boxed{0.0} & 0.3 & 0.7\\
&\area & 3.4 & 2.1 & 1.1 & 0.4 & 0.1 & \boxed{0.0} & 0.3 & 1.0\\
\hline

\multirow{2}*{\makecell{$W$ fusion, $H\rightarrow \mu^-\mu^+$ \\ $\sqrt{s} =360$ GeV } }
&\volume & 7.4 & 1.8 & \boxed{0.0} & 1.9 & 7.3 & 16.4 & 29.2 & 45.7\\
&\area & 9.0 & 2.9 & \boxed{0.1} & 0.9 & 5.2 & 13.6 & 27.0 & 46.4\\
\hline

\multirow{2}*{\makecell{$W$ fusion, $H\rightarrow q\bar{q}$ \\ $\sqrt{s} =360$ GeV } }
&\volume & 3.1 & 1.8 & 0.9 & 0.3 & \boxed{0.0} & 0.0 & 0.3 & 0.8\\
&\area & 3.4 & 2.1 & 1.1 & 0.4 & 0.1 & \boxed{0.0} & 0.3 & 1.0\\
\hline

\multirow{2}*{\makecell{$\mu^\pm$, ${ b\bar{b} \mu \nu_\mu ud }$ \\ $\sqrt{s} =360$ GeV } }
&\volume & 3.7 & \boxed{0.3} & 0.8 & 5.1 & 13.2 & 25.2 & 41.1 & 61.2\\
&\area & 5.0 & 0.8 & \boxed{0.2} & 3.1 & 9.8 & 20.9 & 37.3 & 60.5\\
\hline

\multirow{2}*{\makecell{$ud$, ${ b\bar{b} \mu \nu_\mu ud }$ \\ $\sqrt{s} =360$ GeV } }
&\volume & 2.9 & 1.7 & 0.9 & 0.3 & 0.0 & \boxed{0.0} & 0.3 & 0.7\\
&\area & 3.2 & 2.0 & 1.1 & 0.4 & 0.1 & \boxed{0.0} & 0.3 & 0.9\\
\hline

\multirow{2}*{\makecell{$b\bar{b}$, ${ b\bar{b} \mu \nu_\mu ud }$ \\ $\sqrt{s} =360$ GeV } }
&\volume & 3.1 & 1.9 & 0.9 & 0.3 & 0.0 & \boxed{0.0} & 0.3 & 0.7\\
&\area & 3.4 & 2.1 & 1.1 & 0.4 & 0.1 & \boxed{0.0} & 0.3 & 1.0\\
\hline

\end{tabular}
\end{table}

For the optimal tracker configuration, tracks prefer a smaller tracker radius (in other words, a larger tracker length) compared to the jets for each of these benchmarks.
This is because the resolution of the tracks in the forward region degrades much more than that of the jets.
Therefore, a longer tracker length is required for the tracks to mitigate the poor performance in the endcaps.
The benchmark channels of the $Z$ process prefer a longer tracker compared to those of the $ZH$, $WW$ fusion, and $tt$ processes for both the tracks and the jets.
This is because the particles in the final state of the $Z$-process have a stronger forward distribution; therefore, the resolution in the forward region receives a higher weight.

As for the optimal performance, the optimal average energy resolution for the jets depends mainly on the jet energy, since the dependence of the resolution on the polar angle is small.
The benchmark channel with more energetic jets will have better optimal performance.
For the tracks, both the polar angle distribution and the energy distribution of the tracks affect the optimal average resolution.
We find two simple cases to analyze.
(1) The tracks in the benchmark channels of the $ZH$ and $tt$ processes have a relatively flat distribution in $\cos\theta$.
Therefore, in these two channels, the benchmark channel with more energetic track will have the worse optimal performance.
(2) The optimal average resolution of benchmark channel of $Z$ is comparable to that benchmark channel of the $ZH$.
In the former, the tracks have advantages due to their lower energy and disadvantages due to their more forward direction. Finally, the two factors cancel each other out.

Table~\ref{fig:compare} shows the performance degradation for different tracker radii compared to the optimum for each benchmark channel.
The degradations for jets are generally smaller than those for tracks.
This is simply because the performance of the jets is less sensitive to the polar angle.

\begin{figure}[!hbtp]
\centering

\subfloat[]{\includegraphics[scale=0.73]{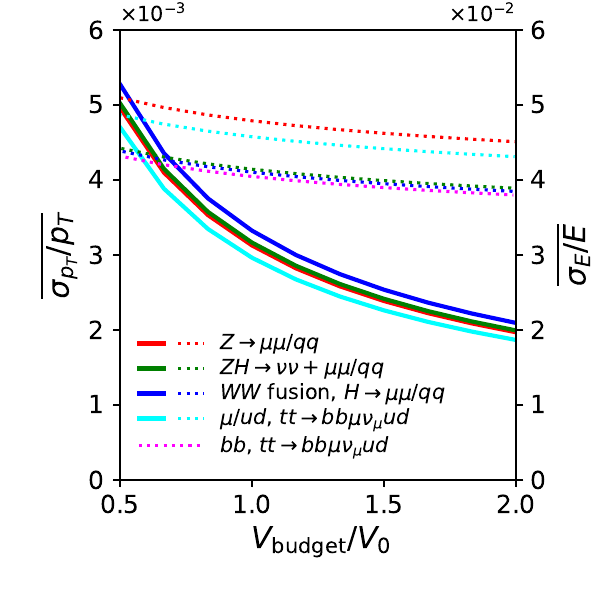} }
\subfloat[]{\includegraphics[scale=0.73]{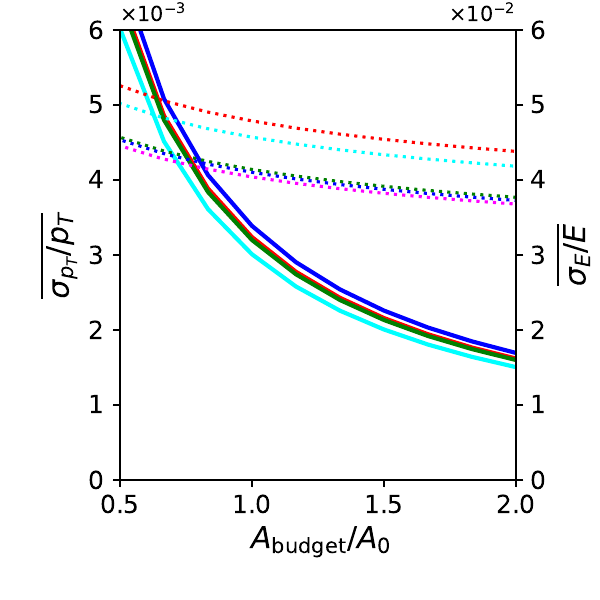} }

\subfloat[]{\includegraphics[scale=0.73]{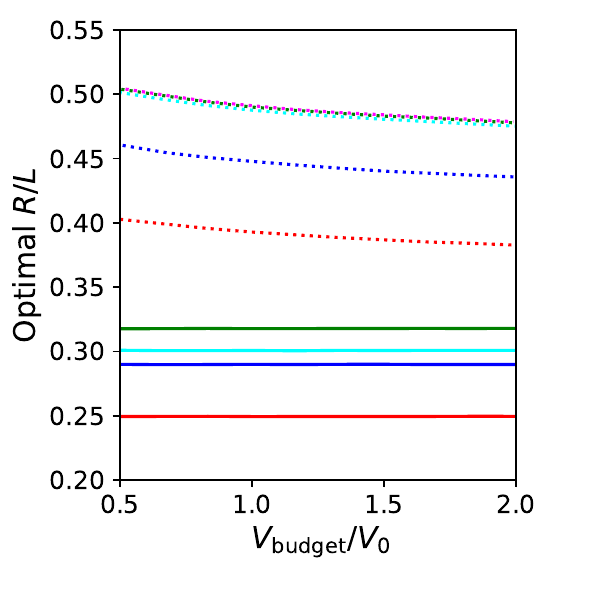} }
\subfloat[]{\includegraphics[scale=0.73]{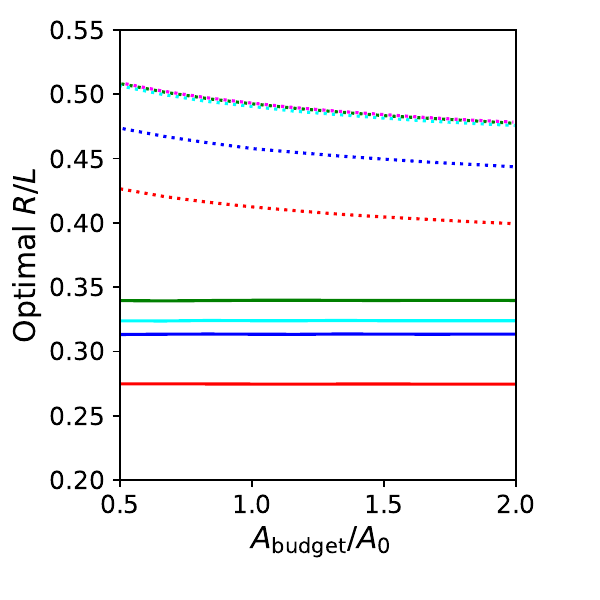} }

\caption{
Top panel: the optimal average resolution as a function of the construction budget.
We use either (a) the tracker volume or (b) the tracker surface area as the cost estimator.
Bottom panel: the optimal $R/L$ as a function of construction budget.
We use either (c) the tracker volume or (d) the tracker surface area as the cost estimator.
The solid line corresponds to the track, and the dotted line corresponds to the jet.
}
\label{fig:RatiovsCost}
\end{figure}

\section{Dependence of the optimal $R/L$ on the construction budget}
\label{sect:variation}

The optimal $R/L$ and corresponding resolutions are determined for different construction budgets, as shown in figure~\ref{fig:RatiovsCost}.
For tracks, the optimal $R/L$ is independent of the construction budget.
When the cost is estimated as the tracker volume, we can rewrite eq.~\eqref{eq:trackTMR} by replacing ($R$, $L$) with ($V$, $\theta^c$):
$$
\sigma_{p_T}/p_T = V^{-2/3} f_V(p_T, \theta;\theta^c),
$$
where $f_V$ is a function of ($p_T$, $\theta$) and
$\theta^c$ is the parameter of the function.
The average resolution would then be $V^{-2/3} \overline{f}_V(\theta^c)$.
The tracker volume as a global scaling factor does not affect the optimal $\theta^c$ and $R/L$.
The optimal average resolution is proportional to $V^{-2/3}$.

When the cost is estimated as tracker surface area, we can reformulate eq.~\eqref{eq:trackTMR} into the form:
$$
\sigma_{p_T}/p_T = A^{-1} f_A(p_T, \theta;\theta^c).
$$
We can draw a similar conclusion:
Fixing $A$ at different values leads to the same optimal $\theta^c$ as well as $R/L$, and the optimal average accuracy is proportional to $A^{-1}$. 
These assertions are confirmed in figure~\ref{fig:RatiovsCost}.

For jets, however, the optimal $R/L$ is only weakly dependent on the construction budget.
As the size of the tracker increases (assuming that the shape remains unchanged),
the performance in the barrel benefits more than that in the endcaps.
In particular, performance in the region near the $z$-axis can't benefit from the longer tracker length.
With a higher construction budget,
it would be better to increase (decrease) the length (radius) to balance the performance in the barrel and endcaps.
Thus, the $R/L$ becomes smaller as the construction budget increases.

\section{Conclusion}
\label{sect:conc}

The CEPC, as a future Higgs and $Z$ boson factory, focuses on testing the Standard Model and searching for new physics.
To fully exploit its physics potential, the optimization of its detector is necessary.
In this paper, we focus on the tracker configuration optimization in terms of the construction cost and two key detector performances, i.e. the track momentum resolution and the jet energy resolution.
Multiple benchmark channels are considered, including $Z\rightarrow f\bar{f}$, $WW$ fusion with $H\rightarrow f\bar{f}$, $ZH\rightarrow\nu\nu f\bar{f}$, and $t\bar{t}\rightarrow b\bar{b}\mu\nu_\mu ud$, covering four operation modes of CEPC with $E_{\rm cm}$ from 91.2\,GeV to 360\,GeV.

\begin{table}[!hbtp]
\centering
\caption{The optimal $R/L$ fixing tracker volume to CEPC baseline.}
\label{tb:RL}

\begin{tabular}{ccccccc}
\hline

$E_{\rm cm}$                & Benchmark                              & Object & Optimal $R/L$   \\
\hline
\multirow{2}*{$91.2\,$GeV}  & \makecell{$Z\rightarrow\mu^-\mu^+$}    & Track & 0.275 \\
                            & \makecell{$Z\rightarrow q\bar{q}$}       & Jet   & 0.412 \\
\hline
\multirow{2}*{$240\,$GeV}   & \makecell{$ZH\rightarrow \nu\nu \mu^-\mu^+$} & Track & 0.339 \\
                            & \makecell{$ZH\rightarrow \nu\nu q\bar{q}$} & Jet & 0.494  \\
\hline
\multirow{2}*{$360\,$GeV}  & \makecell{$WW$ fusion, $H\rightarrow \mu^-\mu^+$} & Track & 0.313 \\
                            & \makecell{$WW$ fusion, $H\rightarrow q\bar{q}$}  & Jet & 0.458 \\
\hline                           
\multirow{3}*{$360\,$GeV} &  \makecell{$\mu^{\pm}$, ${ b\bar{b} \mu \nu_\mu ud }$ } & Track & 0.322 \\
                            & \makecell{$ud$,  ${ b\bar{b} \mu \nu_\mu ud }$}  & Jet & 0.491 \\
                            & \makecell{$b\bar{b}$,  ${ b\bar{b} \mu \nu_\mu ud }$} & Jet & 0.495 \\
\hline                            
\end{tabular}
\end{table}

We first perform a cost constrained optimization of the tracker configuration with the cost fixed to the baseline value.
The cost is estimated by either the tracker volume or the tracker surface area.
In the baseline design of the CEPC detector, the tracker has a cylinder configuration ($R$ = 1.8\,m, $L$ = 4.7\,m) with corresponding volume and surface area of 47.84\,m$^3$ and 73.51\,m$^2$, respectively. The resulting optimal tracker radii range from 1.59\,m to 1.73\,m for track momentum resolution and from 1.82\,m to 1.97\,m for jet energy resolution, depending on benchmark channels.

The optimal $R/L$ fixing tracker volume to CEPC baseline is shown in table~\ref{tb:RL}.
In general, tracks prefer smaller $R/L$ compared to jets for all benchmark channels
because the tracks have a deeper valley in the curve of resolution versus $\cos\theta$, as shown in figure~\ref{PT_resolution_Fix}.
The benchmark channels of the $Z$ process prefer smaller optimal $R/L$ for both tracks and jets compared to those of the $ZH$, $WW$ fusion, and $tt$ processes because the tracks and jets in the $Z$ process have a more forward distribution.
The optimal average resolution for the jets mainly depends on the average energy.
In contract, the optimal average resolution for the tracks depends on not only the average energy but also the polar angle distribution, since the resolution for the track is more sensitive to the polar angle compared to that for the jet.
Compared to the optimal configurations, the degradation of the baseline detector are within 0.4\% for the jets and 12.4\% for the tracks.

Regarding the dependence of the tracker configuration on the construction budget, we find that as the construction budget increases,
the optimal configuration for tracks remains unchanged, and
the corresponding average resolution is inversely proportional to the surface area of the tracker.
For jets, on the other hand, the optimal $R/L$ becomes smaller as the construction budget increases, and the corresponding average resolution improves slightly.
If we double the construction budget, the average resolutions improve by 37\% to 50\% for tracks and 5\% to 9\% for jets, depending on the benchmark channel and the cost estimator.

\section{Acknowledgements}
\label{sect:ackn}

We are grateful to Yuexin Wang for careful reading and constructive suggestions.
This study is supported by the International Partnership Program of Chinese Academy of Sciences (Grant No. 113111KYSB20190030), the Innovative Scientific Program of Institute of High Energy Physics.

\bibliographystyle{JHEP}
\bibliography{ref}

\end{document}